\documentstyle[prl,aps,epsfig]{revtex}
\begin{document}

\def\mev{\hbox{\ MeV}}
 


 \twocolumn[\hsize\textwidth\columnwidth\hsize  
 \csname @twocolumnfalse\endcsname              

\title{Statistical aspects of nuclear coupling to continuum}

\author{S.~Dro\.zd\.z\dag\ddag, J.~Oko{\l}owicz\dag\ddag,
M.~P{\l}oszajczak\ddag~and I.~Rotter\S}
\address{\dag\ Institute of Nuclear Physics, Radzikowskiego 152,
PL - 31342 Krak\'ow, Poland}
\address{\ddag\ Grand Acc\'{e}l\'{e}rateur National d'Ions Lourds,
CEA/DSM -- CNRS/IN2P3, BP 5027, F-14076 Caen Cedex 05, France}
\address{\S\ Max-Planck-Institut f\"ur Physik komplexer Systeme,
D-01187 Dresden, Germany}


\maketitle

\begin{abstract}
Various global characteristics of the coupling between the bound and 
scattering states are explicitly studied based on realistic Shell Model 
Embedded in the Continuum. In particular, such characteristics are related
to those of the scattering ensemble. It is found that in the region of higher
density of states the coupling to continuum is largely consistent with the
statistical model. However,  assumption of channel equivalence in the statistical model 
is, in general, violated.
\end{abstract}

\pacs{PACS number(s): 05.45.+b, 24.60.Ky, 24.60.Lz}

 ]  

\narrowtext

Relating properties of nuclei to the ensembles of random matrices \cite{Brod}
is of great interest. A potential agreement reflects those aspects that are 
generic and thus do not depend on the detailed form of the Hamiltonian matrix, 
while deviations identify certain system-specific, non-random properties of the
system. On the level of bound states the related issues are quite well explored
and documented in the literature \cite{Zele,Dro1}.    
In many cases, however, the nuclear states are embedded in the continuum
and the system should be considered as an open quantum system. 
Applicability of the related scattering ensemble of non-Hermitian random 
matrices \cite{Soko,dtw} has however never been verified by an explicit calculation
due to serious difficulties that such an explicit treatment 
of all elements needed involves. These include a proper handling 
of multi-exciton internal excitations, an appropriate scattering asymptotics
of the states in continuum and a consistent and realistic coupling among 
the two. The recently developed \cite{bnop} advanced computational scheme 
termed the Shell Model Embedded in the Continuum (SMEC) successfully
incorporates such elements and will be used below to study conditions under 
which the statistical description of the continuum coupling applies.
         
Constructing the full SMEC solution consists of three steps. In the first
step, one solves the many-body problem in the subspace $Q$ of (quasi-)bound
states. For
that one solves the multiconfigurational Shell Model (SM) problem :
$H_{QQ}{\Phi}_i = E_i{\Phi}_i$
, where $H_{QQ} \equiv QHQ$ is 
the SM effective Hamiltonian which is appropriate for the SM configuration
space used. For the continuum part (subspace $P$), one solves the coupled 
channel equations :
\begin{eqnarray}
\label{esp}
(E^{(+)} - H_{PP}){\xi}_{E}^{c(+)} \equiv
\sum_{c^{'}}^{}(E^{(+)} - H_{cc^{'}}) {\xi}_E^{c^{'}(+)} = 0 ~  \ ,
\end{eqnarray}
where index $c$ denotes different channels and $H_{PP} \equiv PHP$.
The superscript $(+)$ means that boundary
conditions  for incoming wave in the channel $c$ and
outgoing scattering waves in all channels are used.
The channel states are defined by coupling of one
nucleon in the scattering continuum to the many-body SM state in
$(N - 1)$-nucleus. Finally one solves the system of inhomogeneous coupled 
channel equations :
\begin{eqnarray}
\label{coup}
(E^{(+)} - H_{PP}){\omega}_{i}^{(+)} = H_{PQ}{\Phi}_i \equiv w_i
\end{eqnarray}
with the source term $w_i$ which is primarily given by the
structure of $N$ - particle SM wave function ${\Phi}_i$ and which  
couples the wave function of $N$-nucleon localized states with $(N-1)$-nucleon 
localized states plus one nucleon in the continuum \cite{bnop}. 
These equations define functions ${\omega}_{i}^{(+)}$,  which
describe the decay of quasi-bound state ${\Phi}_i$ in the continuum.

The resulting full solution of SMEC equations is then expressed 
as \cite{bnop,bartz} :
\begin{eqnarray}
\label{eq2}
{\Psi}_{E}^{c} = {\xi}_{E}^{c} + \sum_{i,j}({\Phi}_i + {\omega}_i)
\frac{1}{E - H_{QQ}^{eff}}
\langle{\Phi}_{j}\mid H_{QP} \mid{\xi}_{E}^{c}\rangle \ ,
\end{eqnarray}
where
\begin{eqnarray}
\label{eq2a}
H_{QQ}^{eff} = H_{QQ} + H_{QP}G_{P}^{(+)}H_{PQ} \equiv H_{QQ} + W
\end{eqnarray}
defines the effective Hamiltonian acting in the space of quasibound states.
Its first term reflects the original direct mixing while the second term 
originates from the mixing via the coupling to the continuum.
$G_{P}^{(+)}$ is the Green function for the single particle (s.p.) motion in
the $P$ subspace. This external mixing is thus energy dependent and consists 
of the principal value integral and the residuum :
\begin{eqnarray}
\label{eqpv}
W_{ij}(E) &=&\sum_{c=1}^{\Lambda} \int_{\epsilon_c}^{\infty} dE'
{{\langle\Phi_j \mid H_{QP} \mid \xi_E^c \rangle\langle \xi_E^c \mid H_{PQ} 
\mid \Phi_i \rangle}
\over {E - E'}}\cr 
&-& i\pi \sum_{c=1}^{\Lambda} \langle \Phi_j \mid H_{QP} \mid \xi_E^c \rangle
\langle \xi_E^c \mid H_{PQ} \mid \Phi_i \rangle.
\end{eqnarray}
These two terms prescribe the structure of the real $W^R$ (Hermitian) 
and imaginary $W^I$ (anti-Hermitian) parts of $W$, respectively.
The dyadic product form of the second term allows to express it as
\begin{eqnarray}
\label{eqd}
W^I = - {i \over 2} {\bf V}{\bf V}^T,
\end{eqnarray}
where the $M \times \Lambda$ matrix ${\bf V} \equiv \{ V_i^c \}$ denotes
the amplitudes connecting the state $\Phi_i$ ($i=1,\dots,M$) 
to the reaction channel $c$ ($c=1,\dots,\Lambda $)
\cite{comment}.This form of $W^I$ 
constitutes a starting point towards statistical description of the related
effects. In the latter case one assumes that the internal dynamics is governed
by the Gaussian orthogonal ensemble (GOE) of random matrices. 
Relation of this assumption to the classical chaotic scattering can also
be traced \cite{bdos}.  
The orthogonal invariance arguments then imply that the 
amplitudes $V_i^c$ can
be assumed to be Gaussian distributed and the channels independent \cite{Soko}.
Assuming, as consistent with the statistical ensemble,
the equivalence of the channels 
one then arrives at the following distribution of the off-diagonal
matrix elements of $W^I$ for $\Lambda$ open channels :
\begin{eqnarray}
\label{eqp} 
{\cal P}_\Lambda (W_{ij}^I) = {{\mid W^I_{ij}\mid^{(\Lambda-1)/2} 
K_{(\Lambda-1)/2}(\mid W^I_{ij}\mid)} \over {\Gamma(\Lambda/2)\, \sqrt{\pi} 
\,2^{(\Lambda - 1)/2}}},
\end{eqnarray}
with $\langle{(W^I_{ij})}^2\rangle = \Lambda$. $K_\lambda$ denotes here the 
modified Bessel function.

\begin{figure}[h]
\epsfig{figure=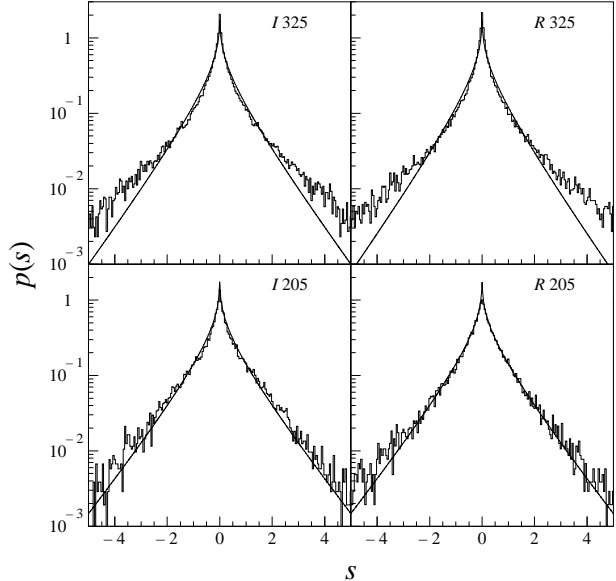,height=8cm}
\caption{Typical projections of distribution of matrix $W$ elements coupling
to the one channel continuum in the SM basis of $J^\pi = 0^{+}, T=0$ states in
$^{24}$Mg (histograms). The projections on imaginary
(left) and real axis (right) are normalized and plotted versus normalized
variable $s = (W_{ij}^X - \langle W_{ij}^X \rangle )/\sigma_X$, where 
$\sigma_X = {\langle {W_{ij}^X}^2\rangle}^{1/2}$, and
$X = I, R$ denotes imaginary and real parts respectively.  In the upper parts
all 325 states were taken into account, while in the lower parts only
205 states in the middle of the spectrum were included.  
The full curves represent $\Lambda = 1$ distribution 
(Eq. (\protect{\ref{eqp}})).}
\label{fig1}
\end{figure}

The physics to be addressed below by making use of the above formalism is that
of a nucleus decaying by the emission of one nucleon. 
As an example, $^{24}$Mg is taken with the inner core of $^{16}$O and the
phenomenological $sd$-shell interaction among valence nucleons \cite{brownwil}.
For the coupling between bound and scattering states a combination of
Wigner and Bartlett forces
is used, with the spin-exchange parameter $\beta = 0.05$ and the overall
strength coupling $V_{12}^{(0)}=650 \mbox{MeV}\cdot\mbox{fm}^3$ \cite{bnop}.
The radial s.p.\ wave functions in the $Q$ subspace
and the scattering wave functions in $P$ subspace
are generated from the average potential of the Woods-Saxon type \cite{bnop}.

In the above SM space, the $^{24}$Mg nucleus 
has 325 $J^{\pi}=0^{+}, T=0$ states. Depending on 
the particle emission threshold, these states can couple to a number of open 
channels. Such channels
correspond to excited states in the neighboring $N-1$ nucleus. 

\begin{figure}[h]
\epsfig{figure=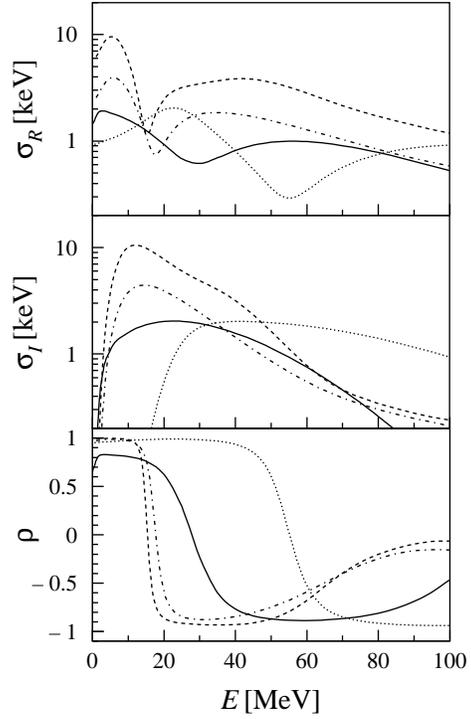,height=10cm}
\caption{Variance of real ($\sigma_R$) and imaginary ($\sigma_I$) 
parts of matrix elements $W_{ij}$
for one open channel and correlation coefficient   
$\rho = (\langle W_{ij}^R \, W_{ij}^I \rangle - \langle W_{ij}^R \rangle 
\langle W_{ij}^I \rangle)/(\sigma_R \, \sigma_I)$
between them. Different line styles 
correspond to different daughter nucleus spins: $1/2$
(full line), $3/2$ (dashed line), $5/2$ (dot-dash) and $7/2$ (dots). All
these quantities are shown as a function of energy of the particle in the 
continuum.}
\label{fig2}
\end{figure}

When testing validity of the statistical model it is instructive to begin
with one open channel and to 
compare the distribution of the corresponding 
matrix elements with the formula (\ref{eqp}) for $\Lambda=1$.
In the example shown in Fig.~\ref{fig1}, the open channel corresponds
to spin 1/2 and its energy to about the middle of the spectrum.
Both the imaginary (left) and real (right) parts of $W$ are displayed. 
The upper part of Fig. \ref{fig1} involves all 325 $J^{\pi}=0^{+}, T=0$ states of 
$^{24}$Mg. Clearly, there are too many large and also too many 
small matrix elements 
as compared to the statistical distribution (solid line) with
$\Lambda=1$ . This may originate from the fact that 
many states in the $Q$ space are localized stronger than allowed by 
the GOE. It is
actually natural to suspect that this may apply to the states
close to both edges of the spectrum. Indeed, by discarding 60 states on both
ends of the spectrum (205 remain), the picture changes significantly
as illustrated in the lower part of Fig. \ref{fig1}. In this case 
the statistical distribution provides a good representation, 
interestingly, also for the real part although applicability of the 
formula (\ref{eqp}) is not directly justifiable as for the imaginary part.
Similar behavior is found for majority of channels except for 
a limited number of them located at the edges of the spectrum.
Hence, the assumption about the Gaussian 
distribution of amplitudes $V^c_i$ is well fulfilled in a generic situation.

\begin{figure}[h]
\epsfig{figure=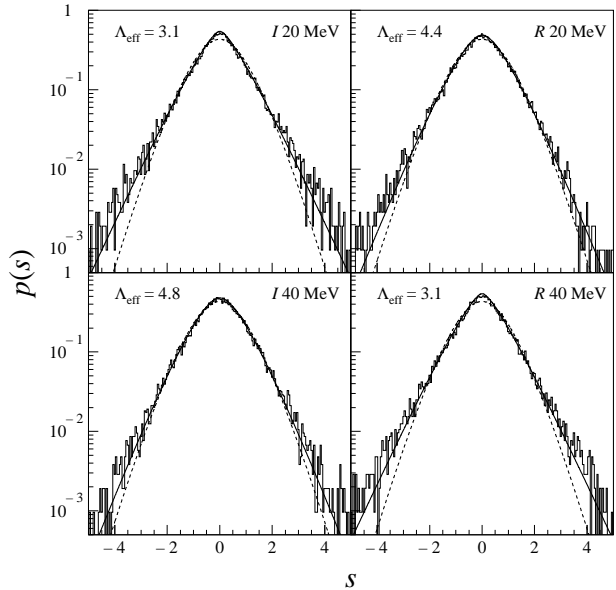,height=8cm}
\caption{The same as lower part of Fig.~\ref{fig1} but for 10 channels of 
spins ranging from
$1/2$ to $7/2$ and two energies of the particle in the continuum (depicted
in the figure). The full curves represent 
${\cal P}_{\Lambda_{\mbox{\tiny eff}}}$ fits with $\Lambda_{\mbox{\tiny eff}}$
indicated while the dashed curves correspond to
 distribution with $\Lambda = 10$.}
\label{fig3}
\end{figure}

As for the equivalence of channels, the conditions are expected to be more 
intricate, especially when different channel quantum numbers are involved,
because the effective coupling
strength depends on those quantum numbers. In addition, such
a coupling strength depends also on energy $E$ of the particle in continuum
so the proportions among the channels may vary with $E$.
This is illustrated in Fig. \ref{fig2} which shows the energy 
dependence of the standard deviations 
of distributions (as in Fig. \ref{fig1}) of relevant 
matrix elements for several different channel spin values, both for real 
and imaginary part of $W$, and their correlation coefficient. 
It should to be noted however that within a given
spin the differences are much smaller.
Instead of trying to identify (with the help of Fig. \ref{fig2}) a sequence of 
$\Lambda$
approximately equivalent channels and to verify the resulting distribution 
of matrix elements of $W$ against formula (\ref{eqp}) we find it more
informative to make a random selection of such channels. An example for
$\Lambda=10$ and two different energies ($E=20$ and $40\mev$) of the particle 
in the continuum is shown in Fig. \ref{fig3}. 
Among these 10 randomly selected channels, two correspond
to spin $1/2$, three to spin $3/2$, three to spin $5/2$ and two to spin $7/2$.
The distributions significantly change as compared to those of the lower
part of Fig. \ref{fig1}. Moreover, ${\cal P}_{\Lambda=10}(W^{I,R}_{ij})$ 
(Eq. (\ref{eqp}) ) (dashed lines) does not provide an optimal representation 
for these explicitly calculated  distributions. 
For $E=20\mev$ particle energy (the upper part of Fig. \ref{fig3}), 
the best fit in terms of the formula (\ref{eqp})
is obtained for $\Lambda_{\mbox{\tiny eff}}=3.1$ for the imaginary part and  
$\Lambda_{\mbox{\tiny eff}}=4.4$ for the real part of $W$. At 
$E=40\mev$ one obtains $\Lambda_{\mbox{\tiny eff}}=4.8$ and 
$\Lambda_{\mbox{\tiny eff}}=3.1$, correspondingly.
This, first of all, indicates that effectively a smaller number of channels
is involved
what is caused by the broadening of the width distribution 
as a result of the non-equivalence of the channels
\cite{broaden}. Secondly, such effective characteristics depend on the energy of 
particle in the continuum,
what in turn is natural in view of the dependences displayed in 
Fig. \ref{fig2}.
It is also interesting to notice that $W^R_{ij}$ obeys functionally similar
distribution as $W^I_{ij}$ although this does not result 
from Eq. (\ref{eqpv}) \cite{comment}. 
 
\begin{figure}[h]
\epsfig{figure=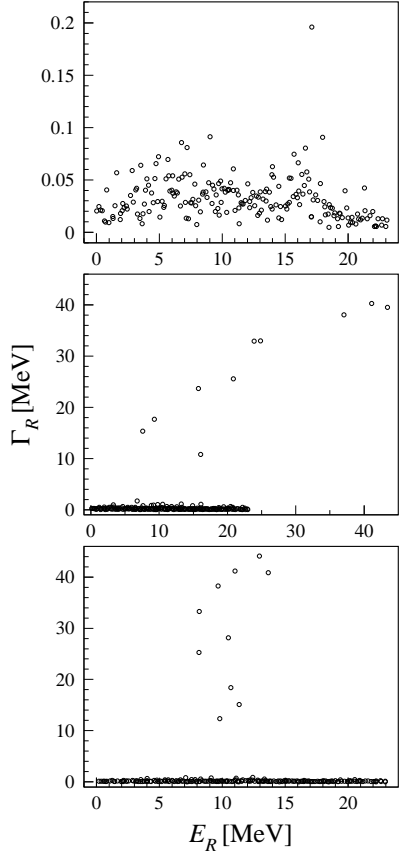,height=11.5cm}
\caption{205 complex eigenvalues for 10 channels and energy of the
particle in the continuum of $40\mev$ are presented as small circles with
coordinates of $E_R$ and $\Gamma_R$. The upper part represents those for
the original residual interaction between $Q$ and $P$ subspaces. 
The middle one is obtained for 7 times stronger interaction 
, and in the lower part
this stronger force is applied to $W^I$ only.}
\label{fig4}
\end{figure}

The fact that generically $\Lambda_{\mbox{\tiny eff}}$ is much smaller than
the actual number of open physical channels can be anticipated from
their obvious non-equivalence in majority of combinations as can be 
concluded from Fig. \ref{fig2}. The global distribution, especially 
in the tails, is dominated by stronger channels.
Due to the separable form of $W$, which
in terms of $\Lambda$ explicitly expresses its reduced dimensionality relative
to $H_{QQ}$, an interesting related effect
in the eigenvalues of $H_{QQ}^{eff}$
 may take place. 
For a sufficiently strong coupling to the continuum one may observe a 
segregation effect among the states, {\it i.e.}, 
$\Lambda$ of them may separate from the remaining $M-\Lambda$ states 
\cite{Rott}. This  effect is especially transparent when looking at the 
structure of $W^I$. For the physical strength $V_{12}^{(0)}$ 
of the residual interaction in
$^{24}\mbox{Mg}$ this effect is negligible, as shown in the upper panel 
of Fig. \ref{fig4}. Only one state  in this case
separates from all others by acquiring a larger width. A 
magnification of the overall strength $V_{12}^{(0)}$ 
of the  coupling to the continuum by a constant
factor $f$ allows further states to consecutively separate. For $f=7$,
all 10 states become unambigously separated as illustrated in the middle 
panel of Fig. \ref{fig4}. Their distance from the remaining, trapped
states reflects approximately the order of their separation when $f$ 
is kept increasing. This nicely
illustrates the degree of non-equivalence of the channels and the fact that
$\Lambda_{\mbox{\tiny eff}} \approx 5$, as consistent with Fig. \ref{fig3} at 
$E=40\mev$, is an
appropriate representation for an effective number of relevant open channels.
It needs also to be noticed that the segregation effect takes place also in the
direction of the real energy axis, though in this sense only three states 
uniquely separate (again consistent with $\Lambda_{\mbox{\tiny eff}}=3.1$ of 
Fig. \ref{fig3}). This direction of the separation originates from the real 
part of $W$. Incorporating an equivalent multiplication factor into $W^I$ 
only, results in a picture as shown in the lower panel of Fig. \ref{fig4}. 
No separation in energy can now be observed anymore.

In summary, the present study indicates that certain characteristics 
of the statistical description of nuclear coupling to the continuum, like the
distribution of coupling matrix elements for one channel continuum, do indeed
apply when the non-generic edge effects are removed. 
On the other hand, in realistic SMEC calculations we find the generic 
nonequivalence of channels which contradicts the orthogonal invariance
arguments and results in strong reduction of the number of effectively  
involved channels.
The quantitative identification and understanding of this effect 
may turn out to be helpful in postulating improved
scattering ensembles which automatically account for this effect, similarly
as various versions of the random matrix ensembles 
invented \cite{Brod,Wong,John} in the context of bound states.
Up to now the statistical models ignore the real part of the matrix connecting
the bound states to the scattering states. The real part of 
$H^{eff}_{QQ}$ is likely to be dominated by $H_{QQ}$, therefore, 
this, in many cases, may be not a bad approximation. 
Keeping in mind a relatively strong energy dependence 
of $W^R$ (see Fig. \ref{fig2}) this however may not be true in some
cases, especially, 
because the segregation of states in energy (along the real
axis) originates from this part. 
Interestingly, $W^R$ is found to obey similar statistical
characteristics as $W^I$. This does 
not however yet mean that the two parts
of $W$ can simply be drawn as independent ensembles. In fact, the 
individual matrix elements $W^I_{ij}$ and $W^R_{ij}$ are often strongly 
correlated and the degree of correlation depends on energy of the particle in 
the continuum.
A more detailed account of such correlations will be presented elsewhere.

We thank K. Bennaceur, E. Caurier, F. Nowacki and M. W\'{o}jcik for useful
discussions.
This work was partly supported by KBN Grant No.~2~P03B~097~16 and by the
Grant No. 76044 of the French-Polish Cooperation.



\begin{references}
\bibitem{comment}
It should be noted that the real part of Eq. (6) cannot be written in the
dyadic product form.
\end{references}
\end{document}